\documentclass{aa}  

\usepackage{graphicx}
\usepackage{txfonts}
\usepackage{xcolor}
\begin{document}

   \title{Gravitoviscous protoplanetary disks with a dust component.}

   \subtitle{III. Evolution of gas, dust, and pebbles}

   \author{Vardan G. Elbakyan
          \inst{1,2}\fnmsep\thanks{vgelbakyan@sfedu.ru},
          Anders Johansen\inst{1}, Michiel Lambrechts\inst{1}, Vitaly Akimkin\inst{3},
          \and
            Eduard I. Vorobyov\inst{2,4,5}
          }

   \institute{Lund Observatory, Department of Astronomy and Theoretical Physics, Lund University, Box 43, 22100 Lund, Sweden
         \and
            Research Institute of Physics, Southern Federal University, Rostov-on-Don, 344090 Russia
        \and
            Institute of Astronomy, Russian Academy of Sciences, Pyatnitskaya str 48, Moscow 119017, Russia
        \and
            University of Vienna, Department of Astrophysics, Vienna, 1180, Austria
        \and
            Ural Federal University, 51 Lenin Str., 620051 Ekaterinburg, Russia
             }

   \date{Received September 15, 2019; accepted November 16, 2019}
   
   \titlerunning{Dust evolution and migration}
   \authorrunning{Elbakyan et al.}

% \abstract{}{}{}{}{} 
% 5 {} token are mandatory
 
  \abstract
   {}
   {We study the dynamics and growth of dust particles in circumstellar disks of different masses that are prone to gravitational instability during the critical first Myr of their evolution. }
   {We solved the hydrodynamics equations for a self-gravitating and viscous circumstellar disk in a thin-disk limit using the FEOSAD numerical hydrodynamics code. The dust component is made up of two different components: micron-sized dust and grown dust of evolving size. For the dust component, we considered the dust coagulation, fragmentation, momentum exchange with the gas, and dust self-gravity.}
   {We found that the micron-sized dust particles grow rapidly in the circumstellar disk, reaching a few cm in size in the inner 100 au of the disk during less than 100 kyr after the disk formation, provided that fragmentation velocity is $30\rm~ms^{-1}$. Due to the accretion of micron dust particles from the surrounding envelope, which serves as a micron dust reservoir, the approximately cm-sized dust particles continue to be present in the disk for more than 900 kyr after the disk formation and maintain a dust-to-gas ratio close to 0.01. We show that a strong correlation exists between the gas and pebble fluxes in the disk.
   We find that radial surface density distribution of pebbles in the disk shows power-law distribution with an index similar to that of the Minimum-mass solar nebula (MMSN) regardless the disk mass. We also show that the gas surface density in our models agrees well with measurements of dust in protoplanetary disks of AS 209, HD 163296, and DoAr 25 systems.
   }
   {Pebbles are formed during the very early stages of protoplanetary disk evolution. They play a crucial role in the planet formation process. Our disc simulations reveal the early onset ($<10^5$ yr) of an inwards-drifting flux of pebble-sized particles that makes up approximately between one hundredth and one tenth of the gas mass flux, which appears consistent with mm-observations of discs. Such a pebble flux would allow for the formation of planetesimals by streaming instability and the early growth of embryos by pebble accretion. We conclude that unlike the more common studies of isolated steady-state protoplanetary disks, more sophisticated global numerical simulations of circumstellar disk formation and evolution, including the pebble formation from the micron dust particles, are needed for performing realistic planet formation studies. 
}


   \maketitle
%
%-------------------------------------------------------------------

%%%%%%%%%%%%%%%%% BODY OF PAPER %%%%%%%%%%%%%%%%%%

\section{Introduction}

Circumstellar disks are the cradles of planet formation. These disks, containing gas and dust particles, are observed to live for up to $\sim$$10$ Myr before being dispersed \citep[e.g.,][]{2011WilliamsCieza}. Planetary systems are assembled from the gas and dust in circumstellar disks. The evolutionary scenario of the disk can have a strong influence on the distribution of the dust content in the disk. Dust plays a critical role not only in the disk evolution and planet formation, but it also determines the observational appearance of the disks. One of the fundamental disk parameters that is important, not only for planet formation, but also for observations, is the dust-to-gas ratio in the disk \citep[e.g.,][]{2007Johansen}. Typically, the constant dust-to-gas ratio is used for disk mass derivation from continuum observations, giving the biggest source of uncertainty in disk mass estimates \citep{2007AndrewsWilliams, 2012Birnstiel}.

However dust and gas are transported in the disk differently, thus leading to the order-of-magnitude local deviations of dust-to-gas ratio in the disk from the canonical 1:100 value \citep{2019VorobyovElbakyan}. Dust particle transport is an essential component of the pebble accretion model, where gas-giant cores are formed from the cm-sized pebbles that have been decoupled from gas \citep{2010Ormel, 2012LambrechtsJohansen}. One of the main problems with planet formation theory is the radial migration of dust particles on timescales that are much shorter than their growth timescales \citep[e.g.,][]{1977Weidenschilling}. The trapping of dust particles at the radial pressure bumps in the disk can prevent the dust particles from carrying out fast inward migration.
%One of the many ideas proposed to prevent the dust particles from fast inward migration is the migration trap, when dust particles are trapped at the radially maximized pressure bump 
\citep{1972Whipple, 2003HaghighipourBoss}. Radial pressure bumps could be formed in different parts of cicumstellar disks - at the snow line, at the dead zone inner edge, and at the disk inner edge \citep{2013Drazkowska, 2014JohansenBlum, 2019Charnoz}. 
%Another possible location of radial pressure bump is the radial distance where gravitational and viscous torques in the inner disk become weak thus forming a "bottle neck" \citep{2019VorobyovSkliarevskii}. Radial pressure bump acts not only as a stopping barrier for inward migrating dust, but also serves as a effective dust growth environment \citep{2012Pinilla}. Ring structures frequently observed in the disks \citep{2018LongPinilla, ??} have been associated with the radial pressure bumps \citep{2019Perez}. Accumulation of dust in the radial pressure bumps can trigger streaming instability (SI), which is a mechanism of rapid planetesimal formation \citep{2005YoudinGoodman, 2007JohansenYoudin}.

Due to the small initial angular momentum or inefficient angular momentum transport in the disk, the resulting disk can become massive and prone to gravitational instability \citep{2005VorobyovBasu, 2006VorobyovBasu, 2010RiceMayo, 2010ZhuHartmann}. Recent observations have shown disks with spiral arms, indicating that they are possibly massive and gravitationally unstable \citep{2016Perez, 2018HuangAndrews, 2018Dullemond}. Gravitational instability and fragmentation of disks are considered as one of the possible mechanisms of giant planet and brown dwarf formation \citep{1998Boss, 2010VorobyovBasu, 2012ZhuHartmann, 2015Stamatellos}. Gaseous clumps forming in the disk serve as a local pressure bumps.  Dust particles accumulated inside the gaseous clumps grow rapidly and possibly form solid cores that could eventually become giant, icy, or terrestrial planets \citep{2010BoleyHayfield, 2013Vorobyov, 2017Nayakshin, 2019VorobyovElbakyan}.

In this paper, we study the evolution and dynamics of dust particles in the circumstellar disk and its surrounding envelope. We use FEOSAD numerical hydrodynamics code that allows us to study the formation and long-term evolution of circumstellar disks. Unlike other studies, where only the evolution of the disk with fixed size dust particles is considered, here we also take into account the dust growth. Moreover, the evolution of disk is considered inside the envelope, self-consistently; thus, the envelope serves as a reservoir of the small dust particles, which are constantly growing and migrating inwards, not allowing the dust in the disk to be exhausted. The studies of disk evolution starting from the pre-stellar core collapse phase are essential for understanding how the building blocks of planets (chondrules, calcium-aluminum-rich inclusions, etc.) are formed \citep{2015Charnoz, 2018Pignatale}. 

The paper is structured as follows: In Section \ref{sec:model}, we introduce the numerical models used. Section \ref{sec:results} presents the main results obtained and in Section \ref{sec:discuss}, we discuss the parameter space study.  Our findings are summarized in Section \ref{sec:concl}. In Appendix~A, we test the dependence of our conclusions on the choice of fragmentation velocity and turbulence strength.

\section{Numerical model}\label{sec:model}
In this section, we briefly describe the numerical hydrodynamic model used in this paper. The formation and evolution of
a star and its circumstellar disk is studied using the FEOSAD code described in detail in \cite{2019VorobyovSkliarevskii}. Here, we only briefly review its main features and
equations. 

Our simulations start from the gravitational collapse of a pre-stellar core of a certain mass, angular momentum, and dust-to-gas ratio and continue into the embedded phase of stellar evolution, during which the protostar and protostellar disk are formed. We introduce a "central smart cell" (CSC) at the coordinate origin with a radius of 1 au to avoid small time-steps and save computational time. The matter is accreted from the CSC on the star with two modes: regular and burst. During the regular accretion mode, it is assumed that the mass accreted from the CSC onto the star is a fraction $\xi$ of the mass accreted from the disk to the CSC. For the CSC we use inflow-outflow boundary condition, in which the matter is allowed to flow freely from the disk to the CSC and vice versa. Thus, the mass accretion rate through the CSC-disk interface ($\dot{M}_{\rm disk}$) can be both positive (when the matter flows from the disk to the CSC) and negative (when the matter flows from the CSC to the disk). The mass accretion rate from the CSC onto the star during the regular accretion mode is calculated as

\begin{equation}
\label{mdot_xi}
\dot{M}_{\rm *,csc} = \begin{cases} \xi \dot{M}_{\rm disk}, & \mbox{if } \dot{M}_{\rm disk}>0 \\ 0, & \mbox{if } \dot{M}_{\rm disk}\leq0 \end{cases}.
\end{equation}
\citet{2019VorobyovSkliarevskii} showed that the mass transport rate inside the CSC affects the evolution of protostellar disks. More specifically, a slow mass transport inside the CSC leads to the formation of more massive, warmer disks with dust particles reaching a decimeter in radius. In all our models, presented in this paper, we use $\xi\approx1.0$, meaning that the mass is transported with a similar efficiency both in the disk and CSC. We plan to consider the mass fluxes in the disk with the different mass transport rates inside the CSC in a follow-up study. More information on burst accretion mode and the model of central smart cell in general can be found in \citet{2019VorobyovSkliarevskii}.

For the outer boundary of the computational domain, the free outflow boundary conditions are imposed so that the matter is allowed to flow out of the computational domain, but is prevented from flowing in. We use the polar coordinates ($r,\phi$) on a two-dimensional numerical grid with $256\times256$ grid zones. The radial grid is logarithmically spaced, while the azimuthal grid is uniformly distributed. The simulations are terminated in the T Tauri phase of disk evolution when the age of the system reaches 1.0 Myr.

The evolution of the disk and the envelope are calculated taking into account the viscous and shock heating, irradiation from the central star and from the background, dust radiative cooling from the disk surface, momentum exchange between gas and dust, self-gravity of gaseous and dusty disks, and turbulent viscosity using the $\alpha$-parametrization \citep{1973ShakuraSunyaev}. The equations of mass, momentum, and energy transport for the gas component are:

\begin{equation}
\label{cont}
\frac{{\partial \Sigma_{\rm g} }}{{\partial t}} + \nabla_p  \cdot ( \Sigma_{\rm g} \mathbf{v}_p ) = 0,  
\end{equation}

\begin{equation}
\label{mom}
\begin{split}
\frac{\partial}{\partial t} \left( \Sigma_{\rm g} \mathbf{v}_p \right) +  \left[\nabla \cdot \left(\Sigma_{\rm g} \mathbf{v}_p \otimes \mathbf{v}_p \right)\right]_p = - \nabla_p {\cal P}  + \Sigma_{\rm g} \, \mathbf{g}_p + \\ + (\nabla \cdot \mathbf{\Pi})_p  - \Sigma_{\rm d,gr} \mathbf{f}_p,
\end{split}
\end{equation}

\begin{equation}
\label{energ}
\frac{\partial e}{\partial t} +\nabla_p \cdot \left( e \mathbf{v}_p \right) = -{\cal P} (\nabla_p \cdot \mathbf{v}_{p}) -\Lambda +\Gamma + \left(\nabla \mathbf{v}\right)_{pp^\prime}:\Pi_{pp^\prime}, 
\end{equation}
where the subscripts $p$ and $p^\prime$ refer to the planar components $(r,\phi)$  in polar coordinates; $\Sigma_{\rm g}$ is the gas surface density;  $e$ is the internal energy per surface area;  ${\cal P}$ is the vertically integrated gas pressure calculated via the ideal  equation of state as ${\cal P}=(\gamma-1) e$ with $\gamma=7/5$; $\mathbf{v}_{p}=v_r \hat{\mathbf r}+ v_\phi \hat{\boldsymbol \phi}$  is the gas velocity in the disk plane; and $\nabla_p=\hat{\mathbf r} \partial / \partial r + \hat{\boldsymbol \phi} r^{-1} \partial / \partial \phi $ is the gradient along the planar coordinates of the disk. The term $\mathbf{f}_p$ is the drag force per unit mass between dust and gas, describing the back-reaction of dust on gas according to the method described in \cite{2018Stoyanovskaya}.

The gravitational acceleration in the disk plane,  $\mathbf{g}_{p}=g_r \hat{\mathbf r} +g_\phi \hat{\boldsymbol \phi}$, takes into account self-gravity of the gaseous and dusty disk components found by solving the Poisson integral \citep[see details in][]{2010VorobyovBasu} and the gravity of the central protostar when formed. Turbulent viscosity is taken into account via the viscous stress tensor  $\mathbf{\Pi}$, the expression for which can be found in \citet{2010VorobyovBasu}. The expressions for the cooling and heating rates $\Lambda$ and $\Gamma$ can be found in \citet{2018VorobyovAkimkin}. We parametrize the turbulent viscosity in the disk using the $\alpha$-prescription of  \citet{1973ShakuraSunyaev} with constant $\alpha$-parameter equal to 0.01, which is consistent with the angular momentum transport rates obtained from the three-dimensional MHD simulations of disks with winds \citep{2009Suzuki}.
%corresponding to the MRI active disk \citep{2018Yang}. 
Models with lower values of the $\alpha$-parameter will be considered in a follow-up study.

Dust in our model consists of two components: small (micron-sized) dust and grown dust with a minimum radius of 1.0~$\mu$m and a variable upper radius of $a_{\rm r}$. The dust size distribution has a fixed power law of $p=-3.5$ in the differential size distribution $dn/da \propto a^p$. All dust in the pre-stellar core is in the form of small dust particles, forming the initial dust mass reservoir. During the core collapse and the disk evolution, the small dust from the reservoir gradually turns into the grown dust. Small dust particles are assumed to be coupled with the gas, while the dynamics of grown dust is controlled by friction with the gas and by the total gravitational potential of the star, and the gaseous and dusty components. At later stages, when the fragmentation becomes important, some mass of grown dust returns to small dust population. The continuity and momentum equations for small and grown dust are defined as:

\begin{equation}
\label{contDsmall}
\frac{{\partial \Sigma_{\rm d,sm} }}{{\partial t}}  + \nabla_p  \cdot \left( \Sigma_{\rm d,sm} \mathbf{v}_p \right) = - S(a_{\rm r}),  
\end{equation}

\begin{equation}
\label{contDlarge}
\frac{{\partial \Sigma_{\rm d,gr} }}{{\partial t}}  + \nabla_p  \cdot \left( \Sigma_{\rm d,gr} \, \mathbf{u}_p \right) = S(a_{\rm r}),  
\end{equation}

\begin{equation}
\label{momDlarge}
\begin{split}
\frac{\partial}{\partial t} \left( \Sigma_{\rm d,gr} \, \mathbf{u}_p \right) +  [\nabla \cdot \left( \Sigma_{\rm d,gr} \, \mathbf{u}_p \otimes \mathbf{u}_p \right)]_p  =  \Sigma_{\rm d,gr} \, \mathbf{g}_p \\ + \Sigma_{\rm d,gr} \mathbf{f}_p + S(a_{\rm r}) \mathbf{v}_p,
\end{split}
\end{equation}
where $\Sigma_{\rm d,sm}$ and $\Sigma_{\rm d,gr}$ are the surface
densities of small and grown dust; $\mathbf{u}_p$ describes the planar components of the grown dust velocity; $S(a_{\rm r}) $ is the rate of dust growth per unit surface area, the expression for which can be found in \citet{2019VorobyovSkliarevskii}; and $a_{\rm r}$ is the maximum radius of grown dust.

The properties of the central protostar (e.g., radius and photospheric luminosity) are calculated using the STELLAR evolution code \citep{2008YorkeBodenheimer,2013HosokawaYorke, 2017VorobyovElbakyan, 2019ElbakyanVorobyov}. The evolution of the central protostar and the circumstellar disk are connected self-consistently. The stellar mass grows according to the mass accretion rate from the disk. The radiative heating of the disk is also calculated self-consistently in accordance with the protostellar photospheric and accretion luminosities.

The initial surface density and angular momentum distributions in our models have the form of:
\begin{equation}
\Sigma_{\rm g}=\frac{r_{0}\Sigma_{\rm g,0}}{\sqrt{r^{2}+r_{0}^{2}}},
\label{eq:sigma}
\end{equation}
\begin{equation}
\Omega=2\Omega_{0}\left(\frac{r_{0}}{r}\right)^{2}\left[\sqrt{1+\left(\frac{r}{r_{0}}\right)^{2}}-1\right],
\label{eq:omega}
\end{equation}
where $\Sigma_{\rm g,0}$ and $\Omega_{0}$ are the angular velocity and
gas surface density at the center of the core; $r_{0}=\sqrt{A}c_{\mathrm{s}}^{2}/\pi G\Sigma_{\rm g,0}$
is the radius of the central plateau, where $c_{\mathrm{s}}$ is the initial isothermal sound speed in the core. Such a radial profile is typical of pre-stellar cores formed as a result of the slow expulsion of magnetic field due to ambipolar diffusion, with the angular momentum remaining constant during axially-symmetric core compression \citep{1997Basu}. The positive density perturbation $A$ is equal to 1.2, making the core unstable to collapse. We consider three numerical models with different total mass of the initial prestellar core. The model parameters are listed in Table \ref{tab:1}. The initial dust-to-gas ratio in all models is 1:100. The initial cores of model L and M have ratio of the rotational to the gravitational energy $\beta=0.24\%$, while the core in model S has $\beta=0.07\%$. Such values are consistent with the observations of pre-stellar cores \citep{2002Caselli}. We use smaller value of $\beta$ in model S to simulate more compact and less massive circumstellar disk. The initial gas temperature in collapsing cores is $T_{\rm init}$ = 20 K.

\begin{table*}
\center
\caption{\label{tab:1}Model parameters. $M_{\mathrm{core}}$ is the initial core mass, $\Omega_{0}$ and $\Sigma_{\rm g,0}$ are the angular velocity and gas surface density at the center of the core, $r_{0}$ is the radius of the central plateau in the initial core, $r_{\mathrm{out}}$ is the initial radius of the core, and $\beta$ is the ratio of rotational to gravitational energy.}
\begin{tabular}{ccccccc}
\hline 
\hline 
Model & $M_{\mathrm{core}}$ &  $\Omega_{0}$ & $r_{0}$ & $\Sigma_{\rm g,0}$ & $r_{\mathrm{out}}$ & $\beta$\tabularnewline
 & [$M_{\odot}$] & [$\mathrm{km\,s^{-1}\,pc^{-1}}$] & [au] & [$\mathrm{g\,cm^{-2}}$] & [au] & [\%] \tabularnewline
\hline 
L & 1.03  & 2.1 & 1200 & 0.20 & 7000 & 0.24  \tabularnewline
M & 0.53  & 4.0 & 617 & 0.38 & 3700 & 0.24 \tabularnewline
S & 0.53  & 2.2 & 617 & 0.38 & 3700 & 0.07 \tabularnewline
\hline 
\end{tabular}
\end{table*}

\section{Results}\label{sec:results}

\subsection{Global evolution}

\begin{figure}
\begin{centering}
\includegraphics[width=1\columnwidth]{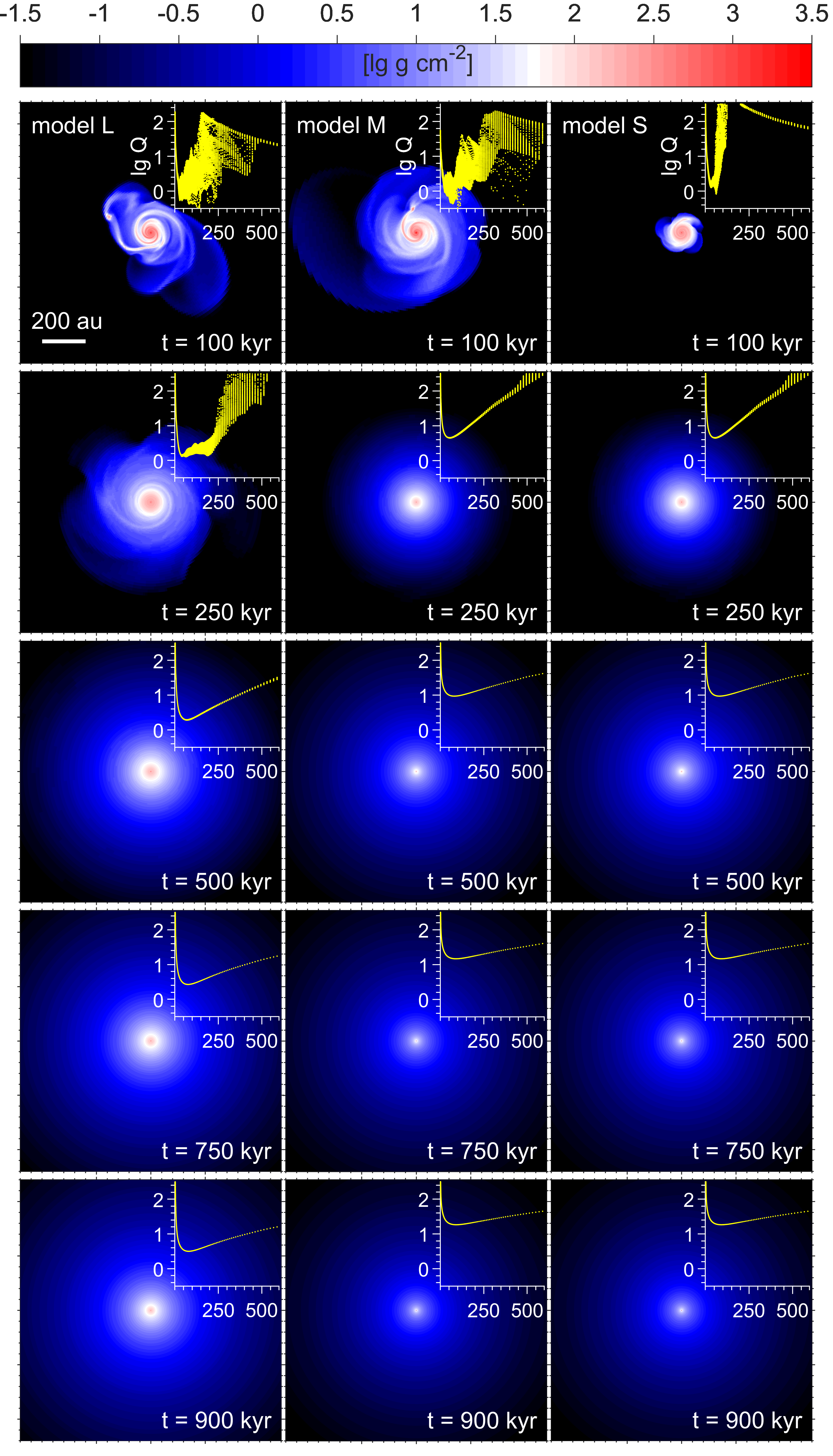}
\par\end{centering}
\caption{\label{fig:1} Gas surface density maps of the inner $1200\times1200$ au$^2$ box at different time moments for all three models. The colorbar is shown in log scale. The insets show the Toomre $Q$-parameter for all azimuthal grid points at a specific radial distance from the star.}
\end{figure}

In this section, we present the global evolution of circumstellar disks in all three models. Figure~\ref{fig:1} presents the gas surface density maps of the inner $1200\times1200$ au$^2$ box for all three models with different initial prestellar core mass at distinct time instances. The computational domain has area $7000\times7000$~au$^2$ (model~L) and $3700\times3700$~au$^2$ (model~M and model~S) including the accreting envelope, but here we show only the inner part with the disk. The time is calculated from the moment of star formation. It is clear from the top row that during its early evolution the disk in model~L has a spiral structure and shows fragmentation, while in the disks of models M and S only spiral structure is visible. To check if the disks fulfill the Toomre gravitational instability and fragmentation criterion \citep{1964Toomre}, in the upper-right corner of each panel we present the Toomre $Q$-parameter for all azimuthal grid points at a specific radial distance from the star. The $Q$-parameter for the mixture of the dust and gas is defined as:
\begin{equation}
Q={\frac{\tilde{c}_{\rm s} \Omega} {\pi G (\Sigma_{\rm g}+\Sigma_{\rm d,sm} + \Sigma_{\rm d,gr})}},
\label{eq:toomre}
\end{equation}
where  $\tilde{c}_{\rm s}=c_{\rm s}/\sqrt{1+\zeta_{\rm d2g}}$ is the modified sound speed \citep{2018VorobyovAkimkin} in the presence of dust and $\zeta_{\rm d2g}$ is the total dust-to-gas mass ratio.

The $Q$-parameter in all our models at $t=100$ kyr shows values less than the threshold value of disk fragmentation ${Q_{\rm fr}=1}$, meaning that the disks in all our models are characterized by gravitational instability. The spiral structures are clearly seen in all the models during the early disk evolution. The $Q$-parameter in all our models grows as the disks evolve. The growth of the $Q$-parameter is due to the decrease of the gas surface density for more than a order of magnitude during initial 900 kyr of disk evolution, while the temperature in the disk during the same time period deceases by only about factor of 3. At $t=250$ kyr the $Q$-parameter in model L becomes greater than the $Q_{\rm fr}=1$, while less than the threshold value for the spiral formation $Q_{\rm sp}=1.5$. We note that precise $Q$ stability value is still under debate and modern numerical simulations show that the disks become unstable and grow non-axisymmetric disturbances, as multi-armed spirals, for $Q<1.5$ \citep[e.g.,][]{2007Durisen,2016Kratter}. 
%The $Q$-parameter in model M and S are higher than $Q_{\rm sp}=1.5$ at $t=250$ kyr, meaning that the disks in these models are stable.
The minimal value of the $Q$-parameter in the model L is higher than $Q_{\rm sp}$ at $t=500$ kyr, meaning that the disk becomes stable. In contrast, the disks in models M and S become stable much earlier and the $Q$-parameter in these models is higher than $Q_{\rm sp}$ already at $t=250$ kyr. The difference in duration of the gravitationally unstable phase of the disks is caused mainly by the difference in the disk masses of our models: the more massive the disk - the longer it will stay unstable. The maximum disk mass in model~L, model~M and model~S are, respectively, 0.3, 0.2 and 0.1~$M_{\odot}$. The subsequent evolution of disks in all models after $t=500$ kyr shows similar behaviour - the disks are gravitationally stable and nearly axisymmetric.

\subsection{Dust growth}

We further analyze the dust growth and the radial distribution of the maximum grain radius $a_{\rm r}$ in the disk. 
%The growth and the dynamics of the dust particles in the disk is controlled mainly with so-called "barriers". 
Due to the aerodynamic drag with the gas, the dust particles lose angular momentum and drift towards the inner parts of the disk, thus experiencing fast radial drift. To grow to the planet embryo size, the dust particles need to overcome this so-called "drift barrier" \citep{1977Weidenschilling}. On the other hand, due to the high relative velocities large dust particles are shuttered rather then stuck together, leading to the so-called "fragmentation barrier" \citep{2008BlumWurm, 2008BrauerDullemond, 2017Gonzalez}. We do not consider the bouncing barrier in this study \citep{2010ZsomOrmel}.

\begin{figure}
\begin{centering}
\includegraphics[width=1\columnwidth]{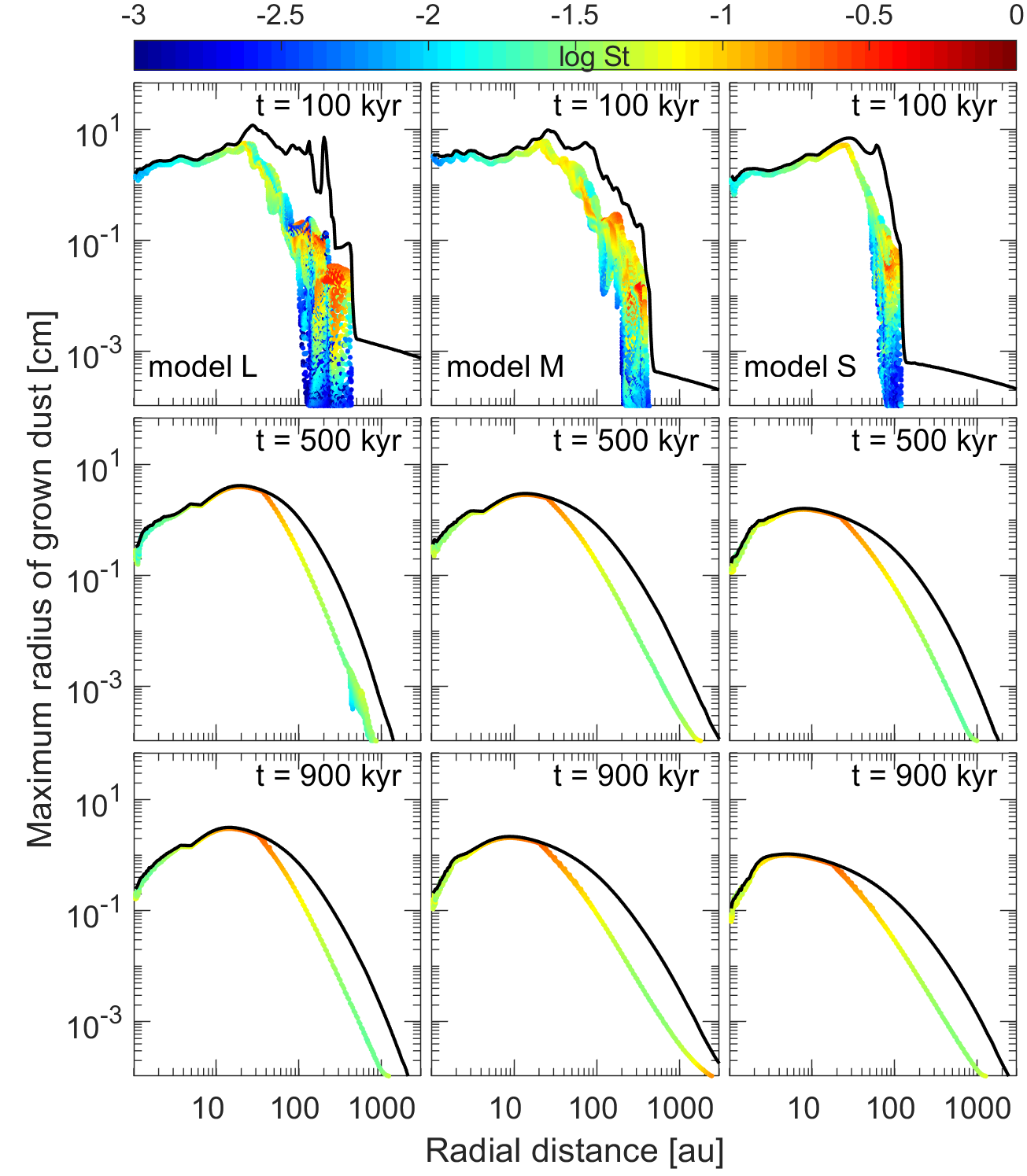}
\par\end{centering}
\caption{\label{fig:4} Radial distribution of the maximum radius of grown dust for all azimuthal grid points shown for consecutive times for all models. Color of dots shows the value of Stokes number for each azimuthal grid point. The colorbar is in log scale. Solid black line marks the dust fragmentation size $a_{\rm frag}$.}
\end{figure}

Figure~\ref{fig:4} shows the maximum dust radius at a given time and orbital distance for all azimuthal grid points in our models. The color of the dots presents the Stokes number ($St$) shown in the colorbar in log scale. The Stokes number, or the dimensionless grain stopping time, is a fundamental parameter controlling the dust dynamics and is defined as:
\begin{equation}
St=\frac{\Omega_{\rm K} \rho_{\rm s} a_{\rm r}}{\rho_{\rm g} c_{\rm s}},
\label{eq:stokes}
\end{equation}
where $\Omega_{\rm K}$ is the Keplerian angular velocity; $\rho_{\rm s}$ = 2.24 g cm$^{-3}$ is the material density of dust grains; $c_{\rm s}$ is the sound speed; $\rho_{\rm g}=\Sigma_{\rm g} / \sqrt{2\pi}H_{\rm g}$ is the gas volume density; $H_{\rm g}$ is the height scale of the gas.

The thick black line shows the fragmentation barrier defined as:
\begin{equation}
\label{afrag}
a_{\rm frag} = \frac{2\Sigma_{\rm g}u^2_{\rm frag}}{3\pi \rho_{\rm s} \alpha c_{\rm s}^2},
\end{equation}
where $u_{\rm frag}$=30~m~s$^{-1}$ is a threshold value for the dust fragmentation velocity \citep[e.g.,][]{2009Wada}. Here we plot the azimuthally averaged value of fragmentation barrier. Given the uncertainties in the $\alpha$-parameter and the fragmentation velocity $u_{\rm frag}$, we discuss the expected outcomes from the changes of these model parameters in the Appendix~A.

Evidently, during less than 100~kyr of initial disk evolution, the main part of the initial micron sized dust in the disks of all our models is efficiently converted into the grown dust particles, reaching few cm in size at the radial distances $\lesssim50$~au. The growth of a dust particle ceases when its radius reaches the fragmentation limit defined with Equation~(\ref{afrag}).
% In the $\alpha=10^{-2}$ model the fragmentation barrier reaches its maximum value at radial distance $r\approx25$~au. 
The size of grown dust particles reaches the fragmentation barrier for the radial distances $\lesssim20$~au, while for the outer parts of the disk it never reaches the fragmentation barrier. As has been shown by \citet{2017Tsukamoto}, the orbital radius
for planetesimal formation by coagulation of fluffy dust particles is $\approx$$20$~au for a gravitationally unstable disk around a solar mass and the dust growth is regulated by the radial drift barrier on radial distance $r\gtrsim20$~au. When dust is transported from regions with larger $a_{\rm frag}$  to regions with smaller $a_{\rm frag}$, we convert some of grown dust mass to small dust to simulate fragmentation. This means that dust growth and dynamics is regulated by the fragmentation barrier in the inner parts of the disk, while in the outer parts of the disk it is regulated by the drift barrier \citep[e.g.,][]{2016Birnstiel}. During the subsequent disk evolution, the fragmentation barrier for all our models shows a well-defined peak at 20~au, decreasing towards the star. This means that the maximum dust radius in the disk is reached at several tens of au.

During the early evolution of the disk, the Stokes number of dust particles in the outer regions of the disks in our models shows high variability, reaching 0.3. Such a high Stokes numbers in the outer part of the disk are reached because the gas density is sufficiently low and the mean free path of gas particles is much higher than the size of dust particles. However, as the disk evolves, the Stokes number in the outer parts of the disk decreases. At the same time the maximum values ($St\approx0.2$) of the Stokes number are reached in the disk at few tens of au, where the dust-size radial distribution shows a peak. The Stokes number of the dust particles, similarly to the size of the particles, is decreasing towards the star for the radial distance of $r<20$ au. 

The envelope and outer parts of the disk, due to the long evolution time scale at large radial distances, play a role of feeding zone for the inner disk, providing not only gas but also small micron-sized dust particles, which grow and migrate inwards, thus compensating the rapid accretion of grown mm- to cm-sized particles onto the central star \citep{2004Kornet, 2007Garaud}. 
%Dust particles with St<<1 are well coupled to the gas, whereas St>>1 are decoupled from it. 
Grown dust particles with mm-cm sizes and relatively high Stokes numbers, also known as pebbles, that are decoupled from the gas, play an important role in formation of planetesimals and planetary cores. \citet{2010JohansenLacerda} showed that protoplanets  can accrete half of their mass from pebbles, thereby giving rise to a planet formation model called pebble accretion \citep{2010Ormel,2012LambrechtsJohansen, 2016Ida, 2017JohansenLambrechts}. Pebble accretion can reduce the planetary core growth timescales below the typical lifetime of the circumstellar disks even at 100 au orbital distances. \citet{2015BitschLambrechts} showed that gas giants formed by pebble accretion are born between 10 and 50 au, and then migrate inward. Although the term "pebble" is widely used, it has no specific definition. Here we define pebbles as dust particles that have the radius greater than $a_{\rm peb}=0.5$~mm. We are not using the Stokes number in the pebble definition, since it implies that the pebble size in the definition will change depending on the radial distance from the star, which in turn will bring a degeneracy in the pebble definition.

\subsection{Dust dynamics}
In this section we study the dynamics of gas and dust in the disk. Much attention is paid to the dynamics of pebbles in the disk, which are essential for the planetesimal formation.

The mass flux in disk at the radial distance $r$ is calculated as:
\begin{equation}
\label{peb_flux}
\dot{M}(r) = \sum_{i=1}^{n}  (M^i/S^i) \mathbf{u}_{\rm r}^i r \Delta\phi,
\end{equation}
where $M^i$ is the mass of matter confined in the $i$th computational cell with surface $S^i$, $\mathbf{u}_{\rm r}^i$ is the radial velocity of the matter in the $i$th computational cell at the radial distance $r$, and $\Delta\phi$ is the spatial grid step in azimuthal direction. Summation is done for the all $n$ azimuthal cells at each radial distance $r$.

Having the minimal radius of pebbles in each cell --  $a_{\rm peb}$, we calculate the mass of pebbles $M_{\rm peb}$ inside each computational cell as:
\begin{equation}
\label{peb_mass}
M_{\rm peb} = \int_{a_{\rm peb}}^{a_{\rm max}} m(a)a^{-3.5}da = \frac{M_{\rm gr.dust} (\sqrt{a_{\rm max}} - \sqrt{a_{\rm peb}})}{\sqrt{a_{\rm max}} - \sqrt{a_{\rm min}}},
\end{equation}
where $a_{\rm max}$ and $a_{\rm min}$ are, respectively, the maximum and the minimum radius of dust particles in the cell, and $M_{\rm tot. dust}$ is the total mass of dust in the cell.

\begin{figure}
\begin{centering}
\includegraphics[width=1\columnwidth]{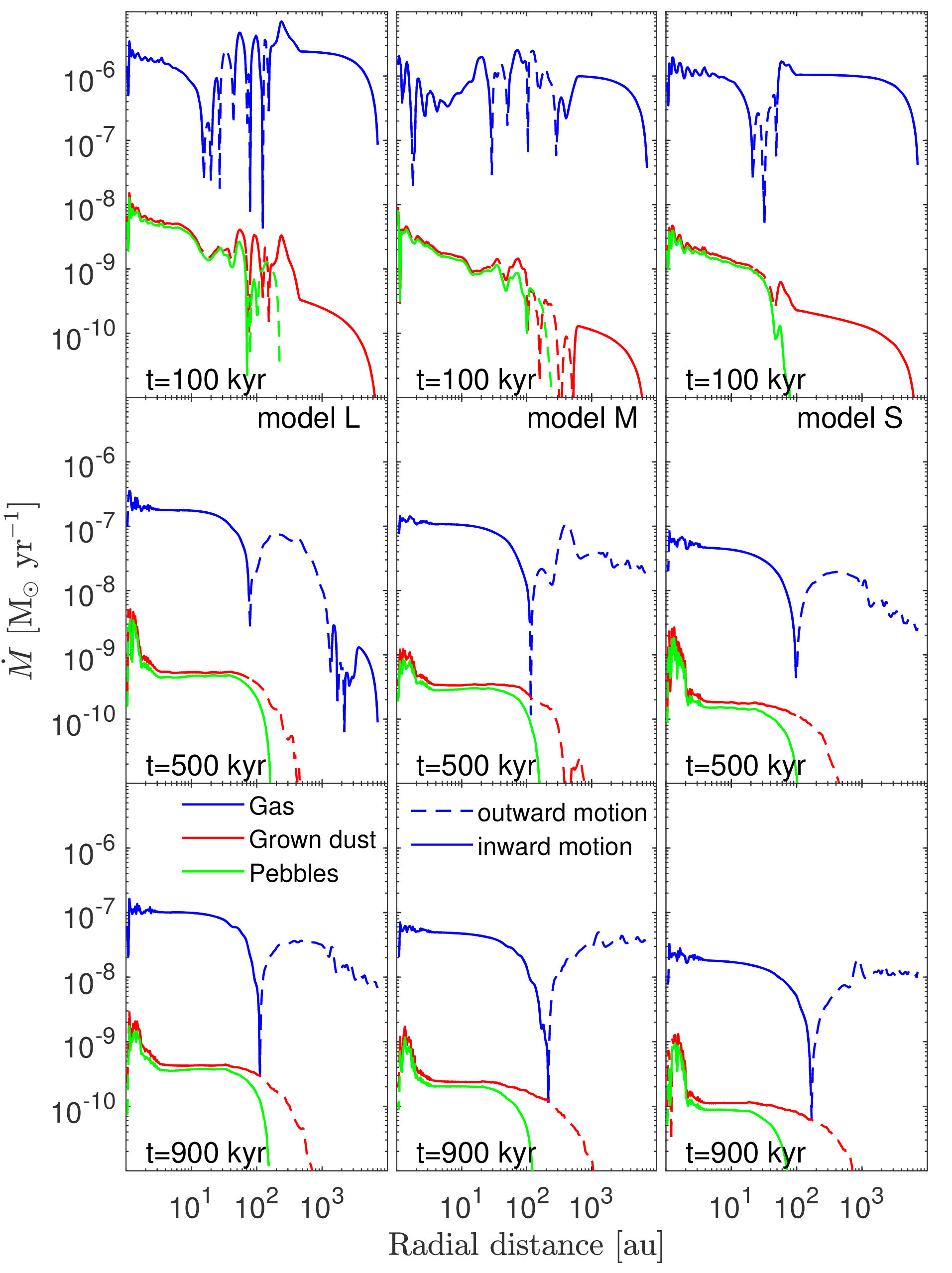}
\par\end{centering}
\caption{\label{fig:5} Azimuthally averaged absolute values of gas (blue curve), grown dust (red curve), and pebble (green curve) accretion (transport) rates versus radial distance from the star at distinct evolutionary times for all our models. Dashed part of the curve shows the outward migration, while the solid part -- the inward migration.}
\end{figure}

It is interesting to see how the gas and dust are accreted on the disk from the envelope and later transported in the disk towards the central star. Figure~\ref{fig:5} shows the absolute values of gas (blue curve), grown dust (red curve), and pebbles (green curve) mass accretion (transport) rates vs radial distance from the central star for different time moments. Dashed part of the curves shows the outward migration, while the solid part shows the inward migration.

During the embedded phase, at $t=100$~kyr, gas accretion from the envelope to the edge of the disk has a quite smooth character for all models, changing from $\dot{M}_{\mathrm g} = \rm few\times10^{-8} \ M_{\odot}yr^{-1}$ at the outer edge of the envelope to $\dot{M}_{\mathrm g} = \rm few \times10^{-6} \ M_{\odot}yr^{-1}$ at the outer edge of the disk. The accretion of grown dust inside the envelope also shows smooth accretion with accretion rates from $\dot{M}_{\mathrm g} = \rm few\times10^{-12} \ M_{\odot}yr^{-1}$ to $\dot{M}_{\mathrm g} = \rm few \times10^{-10} \ M_{\odot}yr^{-1}$. 
%However, the gas and dust transport rates are highly variable inside the disk, changing for more than a order of magnitude.  
In the inner, $r\lesssim20$~au, part of the disk, the gas and the dust during the embedded phase show less variable inward migration, while in the outer parts of the disk ($r\gtrsim20$~au) show highly variable transport rates not only for inward, but also outward migration. The transport rates in the outer region are changing by more than a order of magnitude. Such a behaviour is a result of gravitational instability in the disk during its initial embedded evolution.
The transport rates of pebbles in the inner, $r\lesssim100$~au, part of the embedded disk are similar to the transport rates of grown dust particles, while having slightly lower values.
Transport rates of pebbles in the disk show variations by more than two orders of magnitude. Such a behavior is a result of vigorous gravitational instability taking place in the young embedded disk. The temperature in the inner part of the disk is quite high during the early embedded phase, reaching 150~K (approximately the midplane temperature at the water ice line) at the radial distance $\approx$20~au, while during the later disk evolution 150~K is reached only at the radial distance $\approx$4~au. Similar results are found by \citet{2018Homma} showing that the ice line reaches about 12~au at 0.38~Myr and migrates to 3~au at 1~Myr.

At $t=500$~kyr, when the embedded phase is already over, mass fluxes for both the gas and the grown dust at radial distances $r\gtrsim100$~au become mainly positive, thus showing outward migration. Only the matter from very outer part of computational domain in model L shows inward migration. This is due to the still in-falling very outer parts of the relatively massive initial core. For the inner $r\lesssim100$~au of the disk both the gas and the grown dust show relatively smooth inward migration at about $10^{-7}  \ M_{\odot}\rm yr^{-1}$ and $10^{-9}  \ M_{\odot}\rm yr^{-1}$ rates, respectively. The pebbles also show smooth inward mass transport for $r\lesssim100$~au radial distances having transport rates slightly lower than the transport rates of grown dust particles,
%of few $\times10^{-10}  \ M_{\odot}yr^{-1}$
which are consistent with expected pebble fluxes in the disk \citep{2014LambrechtsJohansen, 2019LambrechtsMorbidelli}. 
We note that the inner $r\lesssim2$~au part of the disk shows some variability connected with the boundary effects between the central smart cell and the active disk. 
At $t=900$~kyr the gas and the grown dust in the outer $r\gtrsim100$~au part of the disk show only outward migration because of the viscous spreading of the disk. The mass fluxes of gas, grown dust and pebbles at the inner $r\lesssim100$~au of the disk show qualitatively similar behaviour as at $t=500$~kyr, but with slightly lower mass transport rates. Note that the pebbles are not forming in the disk on the radial distances $r\gtrsim100$~au, which is consistent with the radial distance of pebble formation line obtained by \citet{2014LambrechtsJohansen}.

\begin{figure}
\begin{centering}
\includegraphics[width=1\columnwidth]{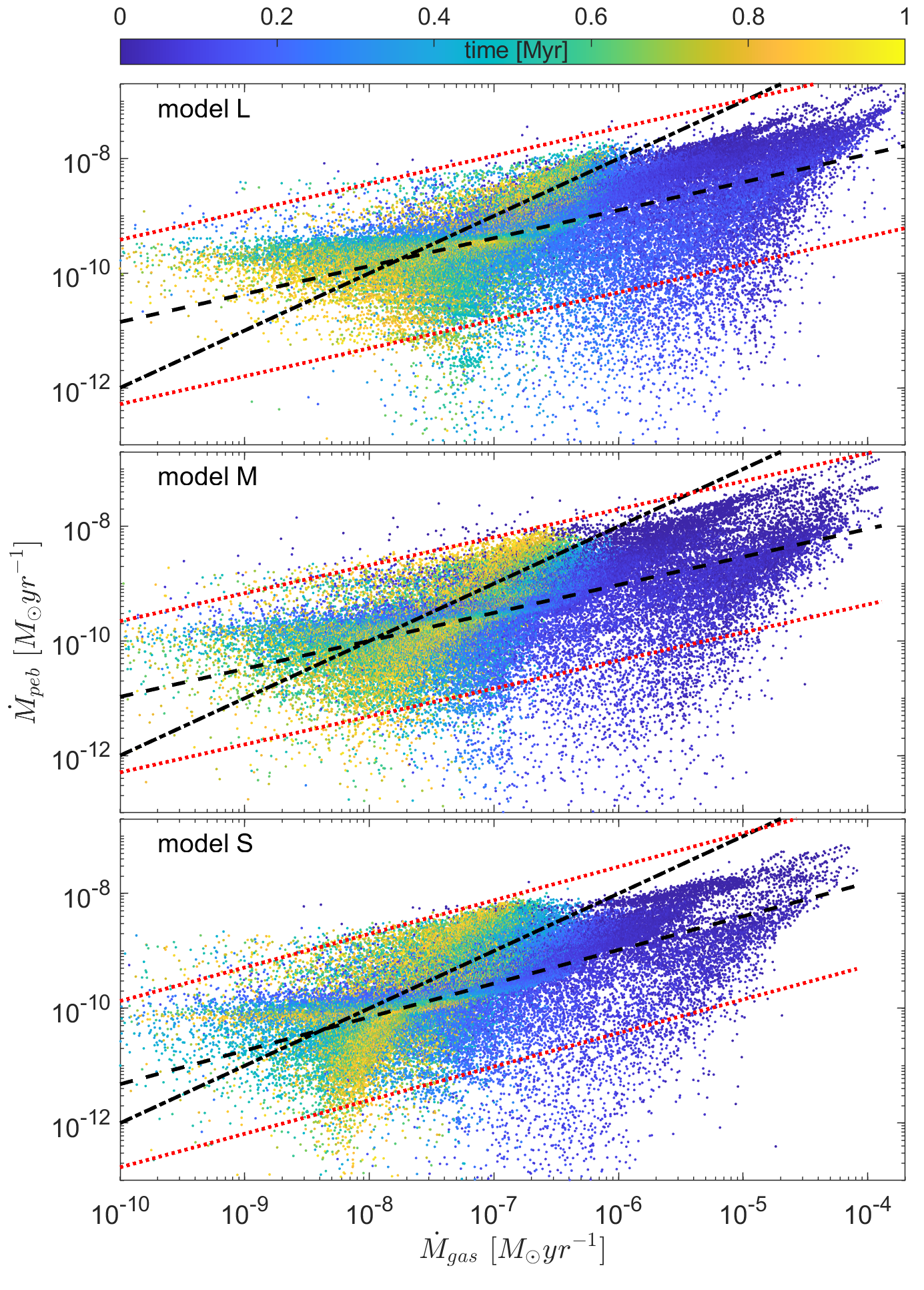}
\par\end{centering}
\caption{\label{fig:11} Relation between pebble mass flux $\dot{M}_{\rm peb}$ and gas mass flux $\dot{M}_{\rm gas}$ for all our models. Color of the dots presents the age of the system in Myr. The black dashed line shows the best-fit curve for each model. The red dotted lines show the $\pm 3\sigma$ deviation from the best-fit values. The dash-dotted line shows the $\dot{M}_{\rm peb}$=$0.01\dot{M}_{\rm g}$ dependence.}
\end{figure}

The pebble mass fluxes in the disk, $\dot{M}_{\rm peb}$, plays an essential role in the formation of rocky Earth-like planets, hot super-Earths and gas giant planets \citep{2019LambrechtsMorbidelli, 2019IzidoroBitsch, 2019BitschIzidoro, 2019JohansenIda, 2019LenzKlahr}. 
An increase in total pebble mass flux in the disk by only a factor of 2 can cause the formation of super-Earth planets instead of Earth-like planets \citep{2019LambrechtsMorbidelli}.
%The increase in total pebble flux in the disk by only a factor of 2 can cause the formation of  gas giant planets instead of super-Earth mass planets \citep{2019IzidoroBitsch}. 
Thus, it is important to understand how the pebbles are transported in the disk. In Figure~\ref{fig:11} we show the pebble mass flux dependence on the gas mass flux. Color of the dots shows the evolutionary time in Myr. The pebble flux shows a tendency to decrease as the disk evolves.  Clearly, the pebble flux in all our models shows a power dependence on gas flux. The dashed line in each panel shows the first order polynomial curve that is the best fit for each model. The $\dot{M}_{\rm gas}$ vs. $\dot{M}_{\rm peb}$ relation can be presented with a polynomial curve as:
\begin{equation}
\label{peb_gas_cor}
\dot{M}_{\rm peb}[M_{\odot} yr^{-1}] = \mathrm{b} \dot{M}_{\rm gas}^\mathrm{a}[M_{\odot} yr^{-1}].
\end{equation}
The polynomial $a$ and $b$ coefficients for all models are presented in Table~\ref{tab:4}. The dotted lines show the $\pm 3\sigma$ deviation from the best-fit values. All the data points that lay between the two red dotted lines are in prediction interval with 99.87\% probability. We overplot the  $\dot{M}_{\rm peb}$=$0.01\dot{M}_{\rm g}$ dependence with the dash-dotted line for the reference.

\begin{figure}
\begin{centering}
\includegraphics[width=1\columnwidth]{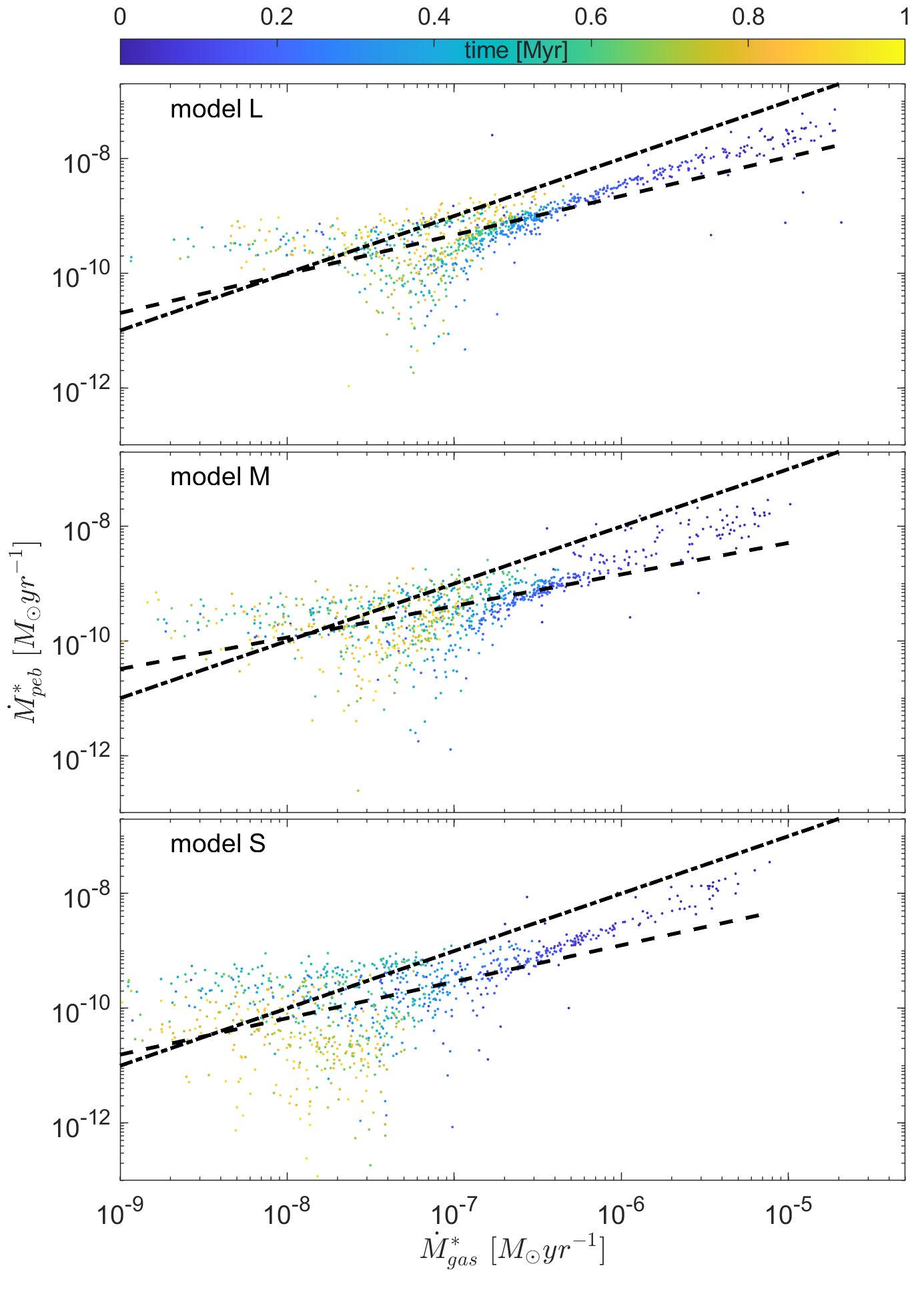}
\par\end{centering}
\caption{\label{fig:12} Similar to Figure \ref{fig:11}, but for the mass accretion rates of gas and pebbles onto the central star.}
\end{figure}

In addition to the transport rates of matter in the disk, it is interesting to study the time dependence of accretion rates of matter from the disk onto the central star. The mass accretion rates could be estimated from the accretion luminosity and the stellar mass and radius (see Equation 8 in \citet{2017VorobyovElbakyan}). The accretion rates are found to correlate with the stellar mass \citep{2004Calvet, 2008HerczegHillenbrand, 2015ManaraTesti} and stellar age \citep{1998HartmannCalvet, 2012Manara, 2014VenutiBouvier}. In Figure~\ref{fig:12}, we present the time dependence of mass accretion rates of gas $\dot{M}_{\rm gas}^{*}$ and pebbles $\dot{M}_{\rm peb}^{*}$ onto the star. The color of dots presents the age of the system in Myr. Evidently, a strong correlation between the gas and pebble mass fluxes exist at the early embedded phase of disk evolution, while during the later disk evolution the correlation becomes weaker. The reason for the weak correlation during the late disk evolution is the fact that as the dust particles grow to the pebble sizes and their Stokes number increases, they decouple from the gas. As a result, the pebble flux in the disk starts to decrease slower than the gas flux and becomes less correlated. We show the best-fit first order polynomial curve for each model with the dashed lines. The polynomial $a$ and $b$ coefficients for each best-fit curve are shown in Table~\ref{tab:5}.

\begin{table}
\center
\caption{\label{tab:4} The $a$ and $b$ parameters for different threshold values of minimum dust size in the pebble definition.  The last row shows the averaged values for all the models.}
\begin{tabular}{ccccccc}
\hline 
\hline 
 & \multicolumn{2}{c|}{0.5 mm} & \multicolumn{2}{c|}{1 mm} & \multicolumn{2}{c}{2 mm}\tabularnewline
 & a & lg(b) & a & lg(b) & a & lg(b)\tabularnewline
\hline 
model L & 0.49 & -5.97 & 0.50 & -5.92 & 0.53 & -5.81 \tabularnewline
model M & 0.49 & -6.09 & 0.51 & -5.99 & 0.53 & -5.96 \tabularnewline
model S & 0.59 & -5.47 & 0.61 & -5.35 & 0.64 & -5.24 \tabularnewline
All models & 0.52 & -5.84 & 0.54 & -5.76 & 0.56 & -5.67 \tabularnewline
\hline 
\end{tabular}
\end{table}

\begin{table}
\center
\caption{\label{tab:5} Similar to Table \ref{tab:4}, but for the mass accretion rates of gas and pebbles onto the central star.}
\begin{tabular}{ccccccc}
\hline 
\hline 
 & \multicolumn{2}{c|}{0.5 mm} & \multicolumn{2}{c|}{1 mm} & \multicolumn{2}{c}{2 mm}\tabularnewline
 & a & lg(b) & a & lg(b) & a & lg(b)\tabularnewline
\hline 
model L & 0.68 & -4.57 & 0.75 & -4.25 & 1.07 & -2.65 \tabularnewline
model M & 0.55 & -5.54 & 0.61 & -5.30 & 0.86 & -3.97 \tabularnewline
model S & 0.64 & -5.09 & 0.76 & -4.35 & 1.00 & -2.98 \tabularnewline
All models & 0.62 & -5.06 & 0.71 & -4.63 & 0.98 & -3.20 \tabularnewline
\hline 
\end{tabular}
\end{table}

\begin{figure}
\begin{centering}
\includegraphics[width=1\columnwidth]{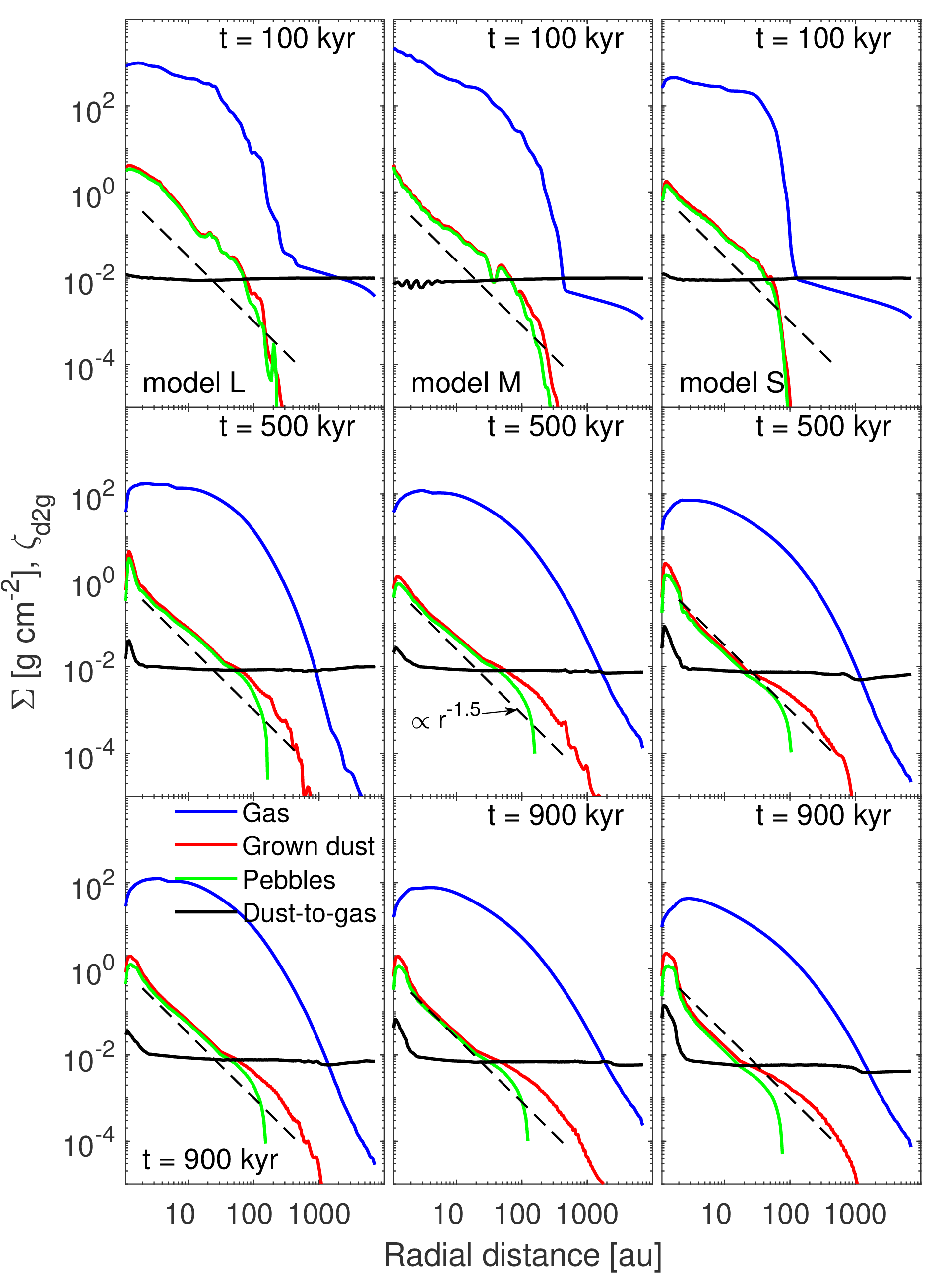}
\par\end{centering}
\caption{\label{fig:9} Azimuthally averaged dust-to-gas mass ratio (black), surface density of gas (blue curve), grown dust (red curve), and pebbles (green curve) versus radial distance from the star at distinct evolutionary times for all our models.}
\end{figure}

Gas surface density distribution in circumstellar disks plays an important role in the process of planet formation. 
Figure~\ref{fig:9} shows the radial distribution of azimuthally averaged dust-to-gas mass ratio ($\zeta_{\rm d2g}$), surface density of gas, grown dust, and pebbles for different time moments. 
The dust-to-gas ratio stays close to the canonical 1:100 value in the entire disk during the disk evolution in all our models, showing increase only in the inner 1~au near the central smart cell. This increase is a result of fast mass transport through the central smart cell leading to the gas depletion near it \citep{2019VorobyovSkliarevskii}.
The radial profile of gas surface density during the early disk evolution is characterized by a slowly decreasing density in the central region ($\lesssim30$ au) and a steeply declining tail at larger radii. 
On the other hand, during the early disk evolution the radial distribution of grown dust and the pebbles have no inner slow-decreasing region, showing steep power law distribution in the inner part of the disk and more steeply decreasing tail at the larger radii. As time progresses, the gas surface density distribution becomes smooth and less steep for the larger radii, while the grown dust and pebble distributions begin to obey a single power-law for the entire disk. The radial profile of pebbles shows $r$-dependency of $\Sigma_{\rm peb}\propto r^{-1.2}$ in model L and M. In model S, the surface density of pebbles becomes proportional to $r^{-1.5}$ after the embedded phase of evolution. Such a dependency is close to the MMSN dependency, which yield $\Sigma_{\rm peb}\propto r^{-1.5}$. Similar results are obtained by \citet{2012Birnstiel} for the dust surface density profile in fragmentation limited regime with relatively large grains present. Our results also agree with the pebble surface densities obtained for constant mass flux and Stokes number for the pebbles \citep{2019LambrechtsMorbidelli}.

\subsection{Comparison with observations}
Here we compare surface densities obtained in our simulations with the surface densities of relatively massive disk in AS 209, HD 163296 and DoAr 25 systems. While neither of the models considered in this paper were specifically tuned for these systems, it is still interesting to check how our models perform against observationally derived disk parameters.
Due to the angular momentum conservation and viscous spreading the disks become larger in time and less dense. Small dust particles that are coupled with the gas will follow the viscous evolution of disk and migrate outwards. Such a distribution of dust in the outer disk complicates the definition of outer edge of the disk, and disk physical and visible sizes for both dust and gas components can differ by a factor of a few \citep{2017Facchini, 2019Rosotti, 2019Trapman}. One of the main disk parameters used in planet formation theory is the disk surface density, which is converted into the total disk mass. However, different methods of disk mass determination are often inconsistent and can vary greatly \citep{2013Bergin}. Recently, \citet{2019Powell} applied the alternative method for disk mass determination suggested by \citet{2017Powell} to the multiwavelength observations of few young disks. The method determines the disk surface density without usage of tracer-to-H$_2$ ratio or an assumed dust opacity model. The disk outer edge is inferred from the millimeter wavelength observations and then related to the maximum radial location of different particle sizes in the disk. The surface density profile of gaseous disk is determined based on the aerodynamic properties of the dust particles. The derived surface density profile is then used as benchmark to scale previously modeled surface density profiles derived from combined multiwavelength dust or CO emission observations.

\begin{figure}
\begin{centering}
\includegraphics[width=1\columnwidth]{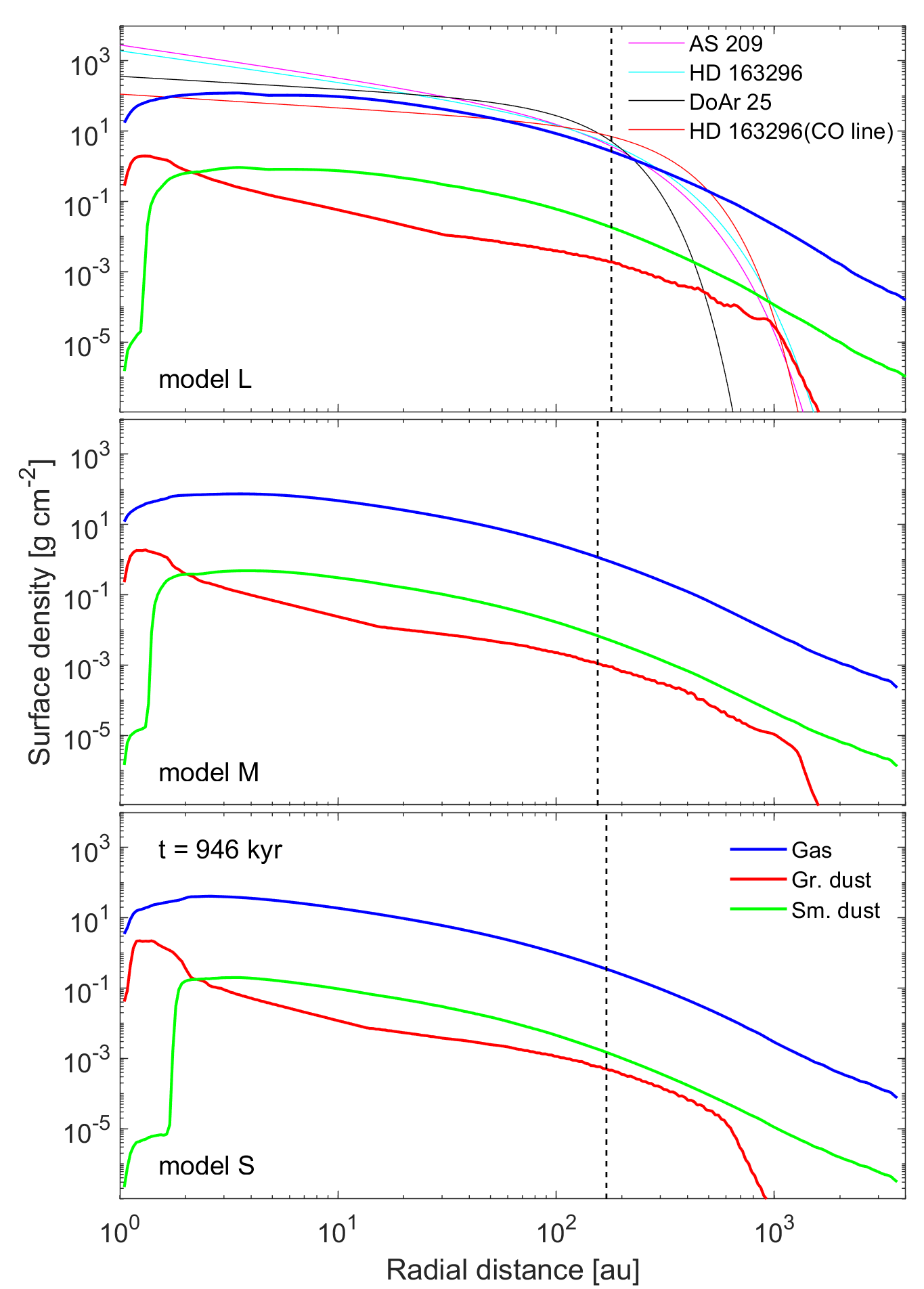}
\par\end{centering}
\caption{\label{fig:8} Radial dependence of the azimuthally averaged surface densities of gas (blue curve), grown dust (red curve), and small dust (green curve) at evolutionary time $t=946$~kyr for all our models. The dashed vertical lines mark the radial distance at which the radius of dust particles $a_{\rm r}=0.85/2\pi$~mm. Thin curves in top panel show the gas surface densities of three observed disks derived by \citet{2019Powell}.}
\end{figure}

In Figure~\ref{fig:8}, we present the azimuthally averaged surface density distributions of gas, grown and small dust at $t=946$ kyr for all our models. The radial distributions show qualitatively same behavior for all our models. The gas surface density slowly decline in the inner few tens of au region and smoothly becomes more steep at larger radii. The radial distribution of small dust surface density for all models follows the shape of gas surface density distribution, except the very inner $2$~au of the disk. Fast mass transport through the central smart cell leads to gas and small dust depletion in the inner 2 au of the disk \citep{2019VorobyovSkliarevskii}. Thus, the inner 2 au of the disks is mainly populated by grown dust particles. The radial distribution of grown dust particles shows different behavior, having a close to a power-law distribution for the entire disk. On the radial distance of about 1000~au the radial distribution of grown dust shows steep decrease in all our models. All the grown dust particles on such high radial distance has drifted inwards. 
In the top panel of Figure~\ref{fig:8}, we overplot the disk surface densities of three\footnote{The surface density for HD 163296 is derived using two different techniques (dust and CO lines) and here we show the both results.} young stellar objects derived by \citet{2019Powell}. The ages and disk masses for the observed objects, along with the disk masses of our models at $t=946$~kyr are presented in Table~\ref{tab:2}. It is assumed that the disk mass is equal to the total mass of gas and dust confined inside the radial distances at which the radial distribution of grown dust shows steep decrease. We compare the observed objects only with our model L because model M and S have much lower masses compared to the disk masses of observed objects. 
%Note that the observed disks are much older then the age of disk in model L. 
The surface densities of observed objects show quite good agreement with gas surface density of model L for radial distances $10\lesssim r \lesssim300$ au. For smaller radial distances ($r\lesssim 10$~au), disk surface densities for DoAr 25 and HD 163296 (CO) show quite good agreement, while for AS 209 and HD 163296 (dust), they overestimate our model results. 

It is important to note that dust surface density distribution is not necessarily mirrored by its thermal emission. The radiation intensity of an isothermal slab of dust with temperature $T$ and absorption coefficient $\kappa_{\nu}$ can be calculated as:
\begin{equation}
I_{\nu}=B_{\nu}(T)(1-\exp\left(-\kappa_{\nu}\Sigma_{\rm d}\right)).
\end{equation}
Here we neglect the scattering, to role of which is actively discussed at the moment \citep{2019ZhuZhang, 2019Liu, 2019CarrascoGonzalez}. To calculate the absorption coefficient $\kappa_{\nu}$ we use Mie theory for a mix of compact carbonaceous and silicate dust as in \citet{2019PavlyuchenkovAkimkin}, assuming their power-law distribution as in our dust evolution model \citep{2018VorobyovAkimkin}. Dust opacity $\kappa_{\nu}$ experiences a distinct maximum near $a_{\rm max}\approx\lambda/2\pi$ \citep{2018BirnstielDullemond} that is reflected as a knee in the intensity radial distribution at the location where grains attain this characteristic size.

Vertical dashed lines in each panel of Figure~\ref{fig:8} mark the radial distance beyond which all the dust particles in the model have radii of $a_{\rm r}<0.85/2\pi$ mm, meaning that these particles will not effectively contribute to the continuum emission at wavelengths $\lambda>0.85$\,mm. The vertical lines for all models lay on the radial distance $168\pm10$~au from the central star, which is consistent with the observational estimates of visible sizes of protoplanetary disks in dust continuum \citep{2018AnsdellWilliams}.

\begin{table}
\center
\caption{\label{tab:2} Disk masses of the observed objects obtained from \citet{2019Powell} along with the disk masses of our models.}
\resizebox{\columnwidth}{!}{\begin{tabular}{ccc|ccc}
\hline 
\hline 
Object & Age &  Disk mass & Object & Age &  Disk mass \tabularnewline
 & [Myr] & [$M_{\odot}$] & & [Myr] & [$M_{\odot}$]  \tabularnewline
\hline 
model L & 0.95  & 0.17 & AS 209 & 1.6  & 0.24    \tabularnewline
model M & 0.95  & 0.06 & DoAr 25 & 2.0  & 0.23 \tabularnewline
model S & 0.95  & 0.02 & HD 163296 & 5  & 0.21 \tabularnewline
& & & HD 163296(CO) & 5  & 0.16 \tabularnewline
\hline 
\end{tabular}}
\end{table}

\begin{figure}
\begin{centering}
\includegraphics[width=1\columnwidth]{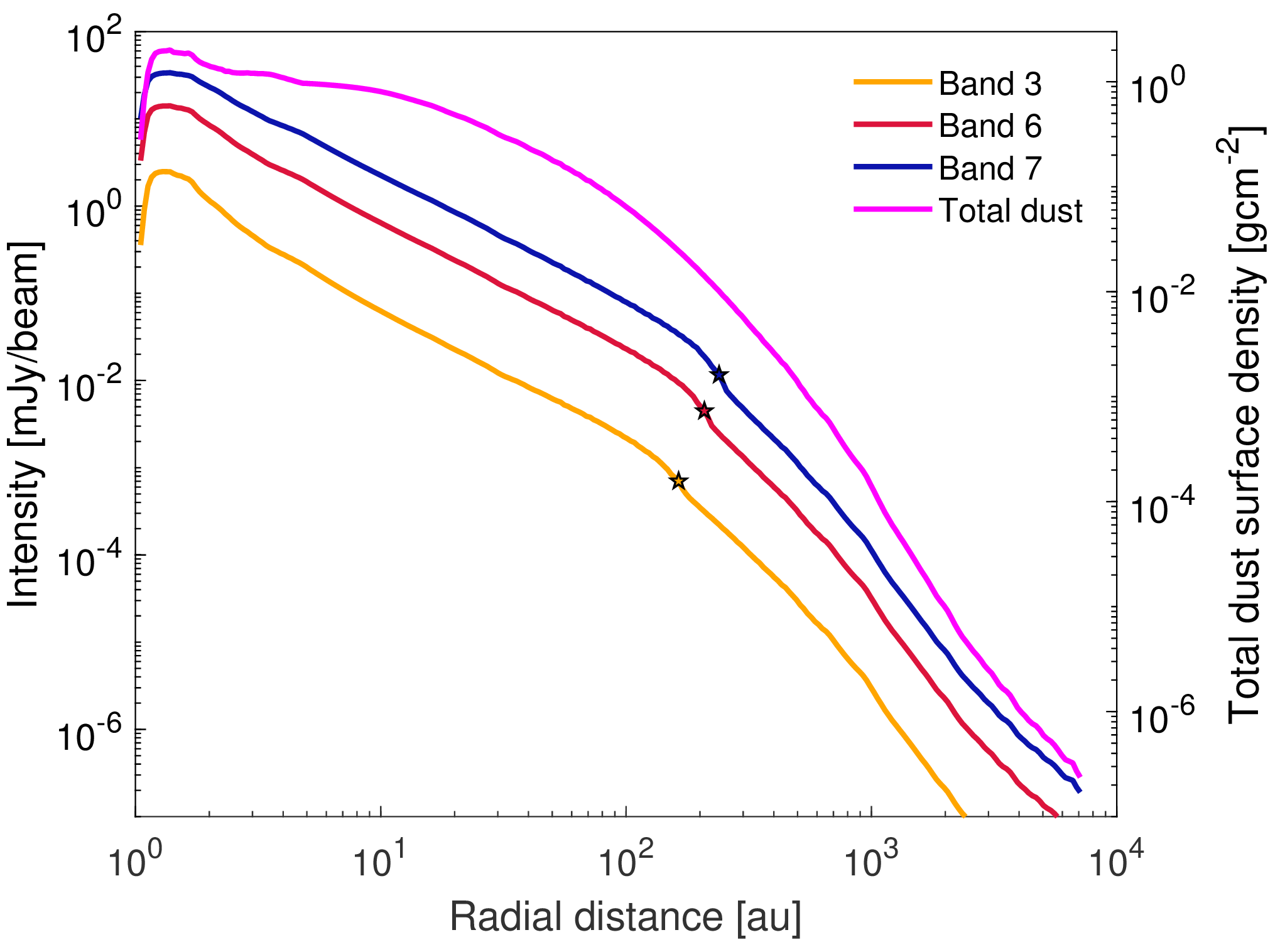}
\par\end{centering}
\caption{\label{fig:18} Synthetic intensities in ALMA Band 3 (2.80 mm), Band 6 (1.30 mm), and Band 7 (0.85 mm) and total dust surface density for model L at $t=964$\,kyr. The stars mark the radial distance at which dust size becomes equal to  $\lambda/2\pi$.}
\end{figure}

Figure~\ref{fig:18} presents the radial distributions of model L synthetic intensities in ALMA Band 3 (2.80 mm), Band 6 (1.30 mm), and Band 7 (0.85 mm). Images are non-convolved, but we present intensities in units per 34\,mas beam. Stars mark the radial distance at which the grown grain size is equal to $\lambda/2\pi$ and the distribution of intensity becomes steeper. We define these radial distances as the outer dust radius for each band, as usually done for the observational data. Figure~\ref{fig:18} shows that the disk emission outside the outer radius traces the density distribution well as the disk periphery is relatively isothermal due to the external heating by interstellar radiation field and the dust opacity coefficient, $\kappa_{\nu}$, also does not vary due to the small size of the grains. The inner disk emission increases much faster towards the star than dust surface density as temperature also increases inwards. We compare the outer dust radii obtained from reproduced intensities of model L with the ones for AS 209, DoAr 25 and HD 163296, taken from \citet{2019Powell}. The results are shown in Table~\ref{tab:3}. The outer dust radius of model L is in quite good agreement with the outer dust radius of DoAr 25, while the ones for As 209 and HD 163296 are overestimated.

\begin{table}
\center
\caption{\label{tab:3} Outer dust radii of the three observed disks for different ALMA bands obtained from \citet{2019Powell} and outer dust radius of our model L obtained from the reproduced intensities.}
\begin{tabular}{cccc}
\hline 
\hline 
  & Band 3 &  Band 6 & Band 7 \tabularnewline
\hline 
model L & 163.7  & 208.5 & 239.4 \tabularnewline
AS 209 & 159  & - & 154.7  \tabularnewline
DoAr 25 & 179.8  & - & 215 \tabularnewline
HD 163296 & - & 101.8 & 121.7 \tabularnewline
\hline 
\end{tabular}
\end{table}

\section{Discussions}\label{sec:discuss}
\subsection{Parameter space study} In order to check how our results for pebble fluxes depend on the threshold value of minimum dust particle radius in pebble definition ($a_{\rm peb}$), we perform a parameter space analysis varying the value of $a_{\rm peb}$. In Table~\ref{tab:4} we present the $a$ and $b$ polynomial coefficients for three different $a_{\rm peb}$ values in pebble definition that are used in Equation~\ref{peb_gas_cor} for all our models. The last row in the Table shows the averaged values for all the models. Clearly, the power-law index for the models with different $a_{\rm peb}$ changes slightly, meaning that the correlation between transport rates of gas and pebbles is weakly dependent on the minimum size of pebbles in range of $0.5-2$~mm.

We also perform parameter space study to check how the mass accretion rates of gas and pebbles onto the star depend on the $a_{\rm peb}$. The $a$ and $b$ polynomial coefficients of best-fit curves for different $a_{\rm peb}$ values are shown in Table~\ref{tab:5}. Unlike the transport rates in the disk, the polynomial coefficients show slight increase for mass accretion rates onto the star. This result is expected because the dust particles are growing during their inward migration in the disk, thus more grown particles with larger sizes are accreted onto the star.

\subsection{Model caveats}
In the current model, we consider only two dust populations of small and grown dust grains with the assumption of a power-law size distribution, which is yet rather simplistic. 
% We plan to implement more than two dust size bins in future studies, as it was done, e.g., in \citet{2010Birnstiel} and more recently in \citet{2019Drazkowska}. 
Multi-grain size simulations of dust growth in hydrodynamically evolving disks have recently become available \citep{2019Drazkowska}, however full-disk simulations with good spatial resolution are very challenging because of high CPU costs. Snow lines play important role in the evolution of the dust particles in protoplanetary disks \citep[e.g.,][]{2019Garate, 2019Musiolik, 2019Ros, 2019Vericel, 2019Ziampras}. In future studies, we plan to improve our model by introducing $\rm H_2O$, $\rm CO$, and $\rm CO_2$ snow lines and adopting dust fragmentation velocity ($u_{\rm frag}$) depending on the position of dust particle relative to the snow line. In addition, other important factors such as bouncing barrier \citep{2010ZsomOrmel} and grain charging \citep{2011Okuzumi, 2015Akimkin, 2019Steinpilz} are planned to be added to the model. It was shown by \citet{2018VorobyovElbakyan} that the numerical resolution becomes important in the study of the migration of dense gaseous clumps. However, we expect that the change in numerical resolution will not dramatically influence the global disk evolution in our model. The model reproduces similar results for gas and dust components in a test problem with a numerical resolutions varying by factor of 2 \citep{2018VorobyovAkimkin}.

\subsection{Planet formation pathways}
The classical planet formation models assume that the dust particles grow up to become kilometer-sized planetesimals, which, in turn, grow into protoplanets colliding and sticking with each other \citep{1972Safronov}. Recently, as an alternative to the classical model, the promising  streaming instability \citep{2005YoudinGoodman, 2015Johansen, 2017Carrera} and the pebble accretion scenarios have been proposed \citep{2012LambrechtsJohansen, 2017JohansenLambrechts}. \citet{2019Tanaka} showed that the pebble accretion could possibly lead to the early planet formation in the Class 0/I disks. Our models showed that only inner $r\lesssim2$~au part of the disk shows dust-to-gas ratios higher than the canonical 1:100 value, which are needed for the streaming instability to take place. Thus, the combination of streaming instability and the pebble accretion could possibly form protoplanetary embryos in the inner, $r\lesssim2$~au, part of the disk. We also predict that the dust-to-gas ratio in the models with lower $\alpha$-parameter will become higher than the canonical value for the radial distances up to 10~au, thus leading to the protoplanetary embryos formation not only in the inner few au of the disk. In our models with $\alpha$$=$$0.01$ the viscous torques are strong in the entire disk, while in the models with the lower $\alpha$-parameter the viscous torques are weak and the matter is transported by the gravitational torques. However, both gravitational and viscous torques are weak in the inner warm region of the disk with low $\alpha$-parameter, leading to the formation of pressure maximum and accumulation of the matter.

\section{Conclusions}\label{sec:concl}

In this paper, we study the long-term evolution of accreting circumstellar disks with their surrounding envelopes using the numerical hydrodynamics code FEOSAD. The joint dynamics of gas and dust (including dust growth) is calculated for about 950 kyr of disk evolution. Three models with different disk masses are considered. The main focus of the study is the evolution and dynamics of grown dust particles, in particular the pebbles.

Our main results can be summarized in the following:

- The dust in the disk grows to the sizes of few cm in the inner ${r\lesssim100}$~au during less than 100 kyrs after the disk formation. Because of a constant supply of small dust from the envelope, the grown few cm sized particles exist in the disk on the radial distances $r\lesssim100$~au for more than 900 kyr.

- The main part of the grown dust particles in the inner ${r\lesssim100}$~au of the disk have pebble sizes. A strong correlation exists between the gas and pebble fluxes in the disks, showing almost universal power-law dependency for disks with different masses. The power-law dependency is more steep for the early embedded phase of disk evolution, while becomes more declivous as the disk evolves. 
%The pebbles are accreted on the central star with accretion rates of few times 1e-10 Msun yr-1, regardless the disk mass.

- The radial distribution of pebbles in the disks shows almost universal power-law distribution, close to the MMSN distribution slope of $\Sigma\propto~r^{-1.5}$.

- The gas surface density of model L shows quite good agreement with the gas surface densities of observed disks obtained from \citet{2019Powell}. The outer dust radius of model L obtained from the reproduced intensities shows a good agreement with the outer dust radius of DoAr 25 obtained from \citet{2019Powell}.

\begin{acknowledgements}
      We thank the anonymous referee for a insightful report, which helped to improve this paper. Research was financially supported by the Ministry of Science and Higher Education of the Russian Federation (State assignment in the field of scientific activity, Southern Federal University, 2020). V.G.E. acknowledges the Swedish Institute for a visitor grant allowing to conduct research at Lund University. AJ was supported by the Swedish Research Council (grant 2018-04867), the Knut and Alice Wallenberg Foundation (grant 2012.0150) and the European Research Council (ERC Consolidator Grant 724687-PLANETESYS). ML was supported by the Knut and Alice Wallenberg Foundation (grant 2012.0150). 
\end{acknowledgements}

\bibliographystyle{aa}
\bibliography{ref_base}

\begin{appendix}
\label{app}
\section{The effect of fragmentation velocity and midplane turbulence on the dust evolution.}

In this appendix, we study how the change in fragmentation velocity $u_{\rm frag}$ and the $\alpha$-parameter in the definition of fragmentation barrier (Equation \ref{afrag}) affects the evolution and the dynamics of dust particles in model~L. We define the modified model~L as model~L2.

The threshold values of fragmentation velocity $u_{\rm frag}$ for different dust particle compositions and sizes have been heavily studied by number of authors either experimentally or numerically \citep{2008BlumWurm, 2009TeiserWurm, 2009Wada, 2010ZsomOrmel, 2013Wada, 2013Meru, 2014Yamamoto, 2015GundlachBlum, 2017BukhariSyed}, giving a wide range of possible values for the velocities. This range varies between 1~m~s$^{-1}$ to 30~m~s$^{-1}$ for the silicate particles, and between 10~m~s$^{-1}$ to 80~m~s$^{-1}$ for icy-grains \citep{2019Charnoz}. In model~L2 we set $u_{\rm frag}=1$~m~s$^{-1}$.

In our models, the same $\alpha$-parameter is used for the description of disk viscosity and the strength of midplane turbulence that affects the fragmentation of dust particles. In model L2, we consider the viscous parameter $\alpha_{\rm v}=10^{-2}$ and the turbulence strength $\alpha_{\rm t}=10^{-4}$. The choice of $\alpha_{\rm t}$ < $\alpha_{\rm v}$ if justified by the models of protoplanetary disks where the midplane is found to be less active \citep[e.g.,][]{2014Turner}.  \citet{2011OkuzumiHirose} found that $\alpha_{\rm t}$ is for about an order of magnitude lower than the $\alpha_{\rm v}$, while based on the observations of the disk around HL Tau, \citet{2016Pinte} found that $\alpha_{\rm t}$ is of the order of a few~$10^{-4}$. In model~L2 in Equation \ref{afrag} we set $\alpha=\alpha_{\rm t} = 10^{-4}$, which is also consistent with the values of $\alpha_{\rm t}$ used in other protoplanetary disk models \citep[e.g.,][]{2017Carrera,2018Drazkowska}.

\begin{figure}
\begin{centering}
\includegraphics[width=1\columnwidth]{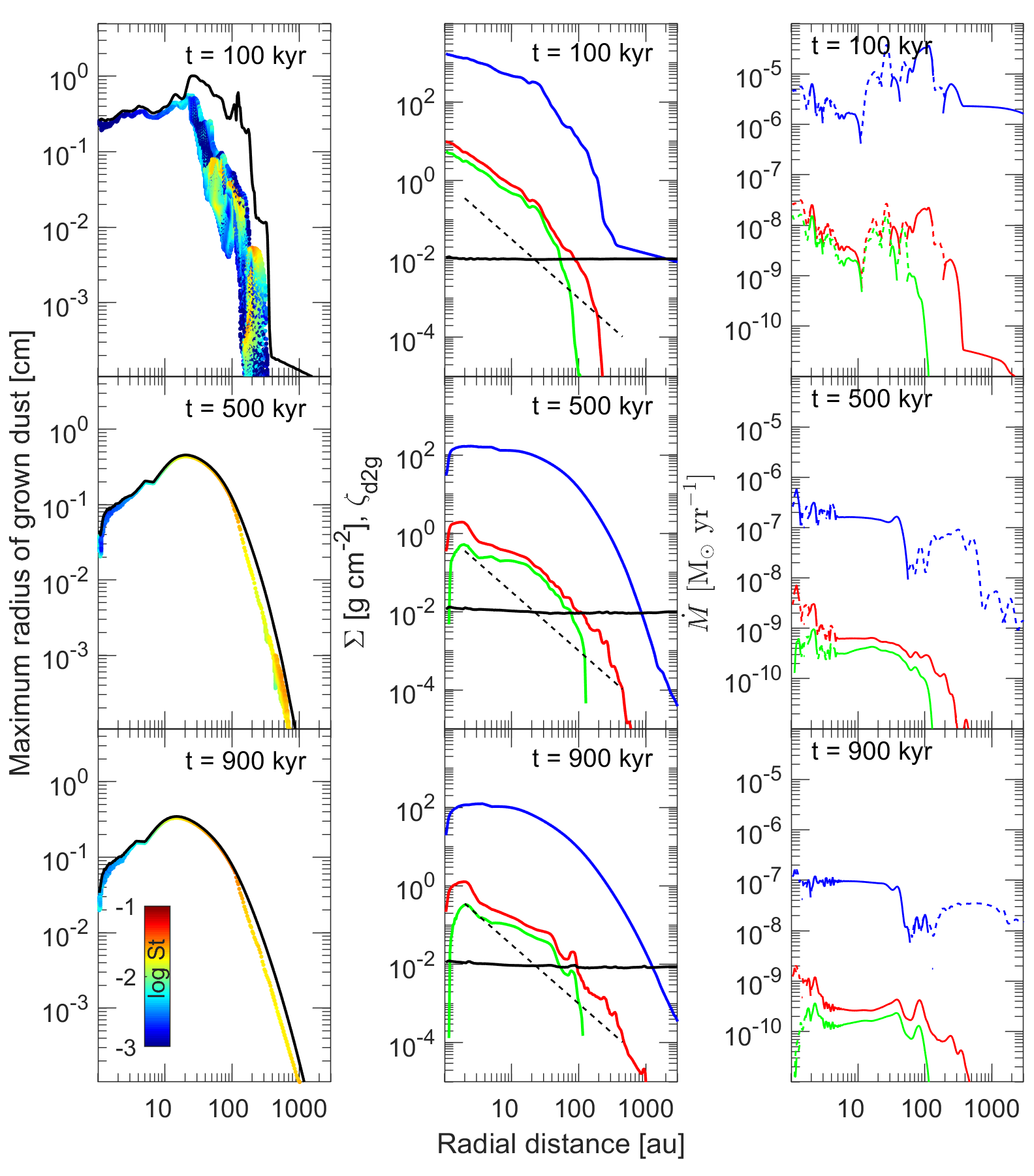}
\par\end{centering}
\caption{\label{fig:13} {\bf Left column:} Radial distribution of the maximum radius of grown dust for all azimuthal grid points in model~L2. Color of dots shows the value of Stokes number for each azimuthal grid point. Solid black line shows the dust fragmentation size $a_{\rm frag}$. {\bf Middle column:} Azimuthally averaged dust-to-gas mass ratio (black), surface density of gas (blue curve), grown dust (red curve), and pebbles (green curve) vs. radial distance from the star. The dashed line shows the power-law slope $\propto r^{-1.5}$. {\bf Right column} Azimuthally averaged absolute values of gas (blue curve), grown dust (red curve), and pebble (green curve) accretion (transport) rates vs. radial distance from the star. Dashed part of the curve shows the outward migration, solid part - the inward migration.}
\end{figure}

The left column in Figure~\ref{fig:13} shows the radial distribution of maximum grown dust radius for all azimuthal grid points at distinct time moments. The Stokes number is shown with the color. Solid black line marks the dust fragmentation size. Unlike model L, the maximum radius of dust particles in model~L2 reaches only few mm. However, the evolution of dust in model~L2 is qualitatively similar to the one in model~L. The dust evolution is regulated with the drift for the radial distance $r\gtrsim20$~au, while with the fragmentation for the radial distance $r\lesssim20$~au. As a result, a well-defined peak in the maximum dust radius is developed at $\approx20$~au. The fragmentation radius decreases closer to the star, reaching few~$\times10^{-2}$~cm. After the end of embedded phase, the fragmentation radius at $r=1$~au becomes lower than 0.5~mm (the minimum pebble size), meaning that no pebbles (dust particles greater than 0.5~mm in radius) are accreted onto the star.
%For evolutionary times after the end of embedded phase, the fragmentation radius at $r=1$~au becomes lower than the minimum pebble size of 0.5~mm, meaning that no pebbles are accreted onto the star. 
As a result of lower maximum dust radii, the Stokes number of dust particles in model~L2 is about an order of magnitude lower than its counterpart in model~L. The Stokes number of dust particles in the inner 10~au of the disk is of order $10^{-3}$, meaning that the dust particles are well coupled with the gas.

The middle column in Figure~\ref{fig:13} shows the radial distribution of azimuthally averaged dust-to-gas mass ratio (black), surface density of gas (blue curve), grown dust (red curve), and pebbles (green curve) at distinct evolutionary times. The power-law slope $\propto r^{-1.5}$ is shown with the dashed line. The evolution of gas surface density in model~L2 is similar to the one in model~L, while the evolution of grown dust and pebble surface density differs. It is easy to notice that unlike model~L, in model~L2 the grown dust and pebble surface density for the radial distance $r\lesssim20$~au does not follow the $r^{-1.5}$ slope and becomes more shallow. This is a result of lower fragmentation radius for the inner part of the disk of model~L2 compared to the model~L. We note that the surface density of pebbles sharply drops at radial distance $r\approx1.1$~au because at this distance the fragmentation radius becomes lower than the minimum radius of pebbles. The azimuthally averaged dust-to-gas ratio ($\zeta_{\rm d2g}$) in model~L2 stays equal to the initial $10^{-2}$ value for the entire disk at all evolutionary time moments.

The right column in Figure~\ref{fig:13} shows the radial distribution of azimuthally averaged absolute values of gas (blue curve), grown dust (red curve), and pebble (green curve) accretion (transport) rates at distinct evolutionary times. Dashed line shows the outward migration, while the solid line shows the inward migration. At the early embedded phase the transport rates of gas, grown dust, and pebbles in model~L2 similar to model~L show high variability with inward and outward migration. The values of transport rates in model~L2 are of the same order as in model~L, differing only by factor of few. At the evolutionary times after the end of embedded phase, transport rates of gas, grown dust and pebbles show relatively smooth inward migration at the radial distance $r\lesssim100$~au with values similar to the ones in model~L. In contrast to model~L, the transport rate of pebbles rapidly decreases at $r\approx1.1$~au.

\begin{figure}
\begin{centering}
\includegraphics[width=1\columnwidth]{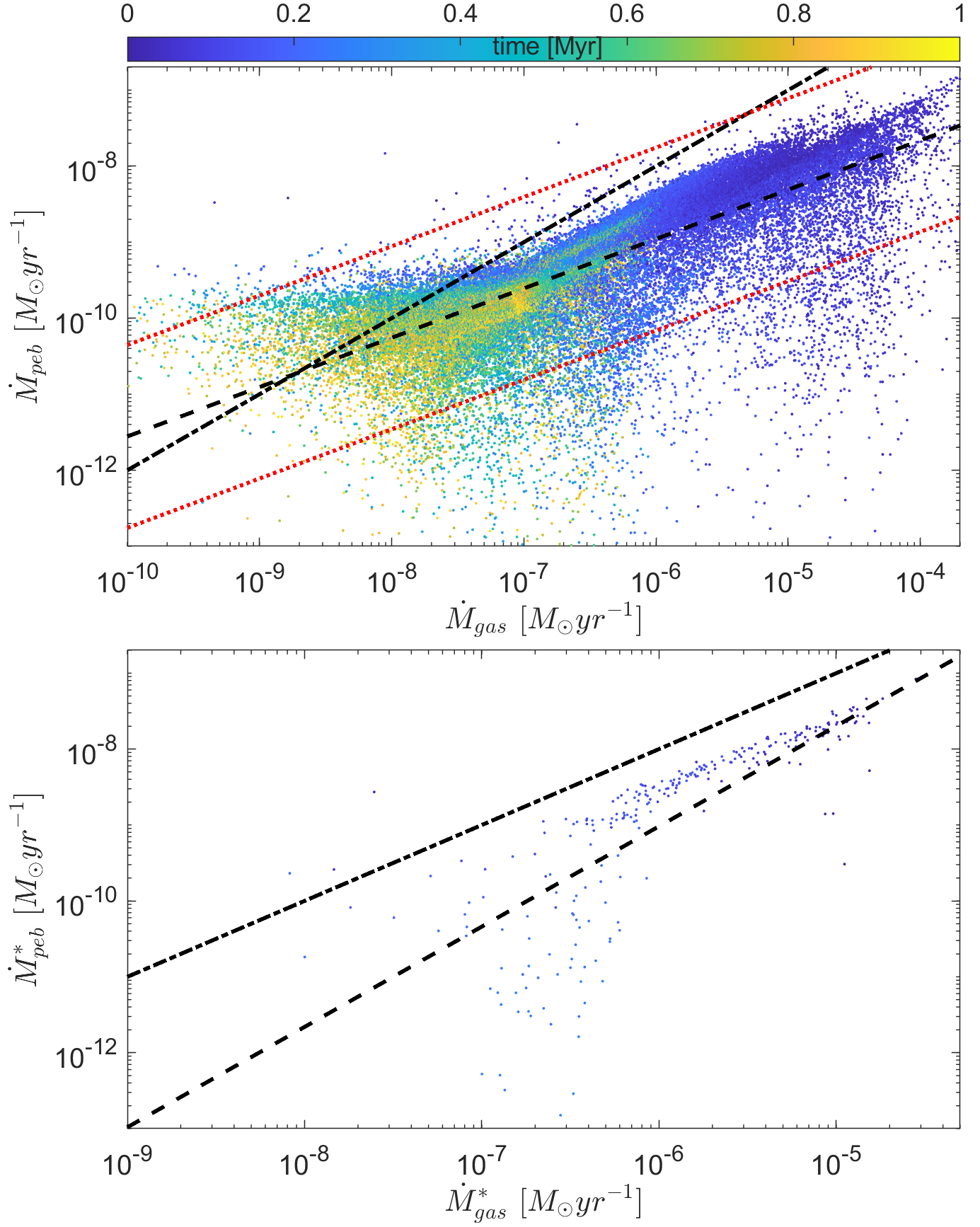}
\par\end{centering}
\caption{\label{fig:14} Relation between pebble mass flux $\dot{M}_{\rm peb}$ and gas mass flux $\dot{M}_{\rm gas}$ in the entire disk (top panel) and onto the central star (bottom panel) for model L2. Color of dots presents the age of the system in Myr. The black dashed line shows the best-fit curve for the data in each panel. The red dotted lines show the $\pm 3\sigma$ deviation from the best-fit values. The dash-dotted line shows the $\dot{M}^*_{\rm peb}$=$0.01\dot{M}^*_{\rm g}$ dependence.}
\end{figure}

We also check how the pebble and gas mass fluxes in the disk are affected by the change of the parameters in the fragmentation radius definition. The top panel of Figure~\ref{fig:14} presents the dependence of the pebble mass flux $\dot{M}_{\rm peb}$ on the gas mass flux $\dot{M}_{\rm gas}$ in the disk. The color of the dots represent the age of the system. Similarly to model L, there is a strong correlation between the pebble and gas fluxes in model L2. The polynomial coefficients $a$ and $b$ of the best-fit curve for $\dot{M}_{\rm gas}$ vs. $\dot{M}_{\rm peb}$ relation are presented in the first row of Table~\ref{tab:6}. The polynomial coefficients in model~L2 are close to the ones in model~L, but slightly higher.
The bottom panel of Figure~\ref{fig:14} shows the time dependent relation between the mass accretion rates of gas $\dot{M}_{\rm gas}^{*}$ and pebbles $\dot{M}_{\rm peb}^{*}$ onto the central star in model~L2. We note that the lower values of $\dot{M}_{\rm peb}^*$ are missing due to the fact that the pebbles are absent at the radial distance $r\lesssim1.1$~au after the end of embedded phase. The polynomial coefficients $a$ and $b$ of the best-fit curve for $\dot{M}_{\rm gas}^{*}$ vs. $\dot{M}_{\rm peb}^{*}$ relation are presented in the second row of Table~\ref{tab:6}. The coefficients in model~L2 are slightly higher than in model~L for the pebbles sizes of 0.5 and 1~mm, while slightly lower for the pebble size of 2~mm. 

\begin{table}
\center
\caption{\label{tab:6} Parameters $a$ and $b$ for different threshold values of minimum dust size in the pebble definition. The top row shows the values for the gas and pebble fluxes in the disk, while the bottom row shows the values for the gas and pebble accretion rates onto the star.}
\begin{tabular}{ccccccc}
\hline 
\hline 
model L2 & \multicolumn{2}{c|}{0.5 mm} & \multicolumn{2}{c|}{1 mm} & \multicolumn{2}{c}{2 mm}\tabularnewline
 & a & lg(b) & a & lg(b) & a & lg(b)\tabularnewline
\hline 
Disk & 0.65 & -5.06 & 0.71 & -4.83 & 0.75 & -4.89 \tabularnewline
Star & 1.32 & -1.09 & 1.06 & -2.60 & 0.81 & -4.04 \tabularnewline
\hline 
\end{tabular}
\end{table}

\end{appendix}

\end{document}